\documentclass[a4paper,12pt]{article}

\def\displayandname#1{\rlap{$\displaystyle\csname #1\endcsname$}%
                      \qquad \texttt{\char92 #1}}

\usepackage{epsfig}

\begin{document}

\title{Violations of Einstein's time dilation formula in particle decays}

\author{Eugene V. Stefanovich \and \small \emph{2255 Showers Dr., Apt. 153, 
Mountain View, CA 94040, USA} \and \small $eugene\_stefanovich@usa.net$}

\maketitle

\begin{abstract}
A rigorous quantum relativistic approach has been used to calculate the
relationship between the decay laws of an unstable particle seen from
two inertial frames moving with respect to each other. In agreement
with experiment, it is found that
the usual Einstein's time dilation formula is rather accurate in this
case.
However, small corrections to this formula were also obtained. Although the
observation of these corrections
is beyond the resolution of modern experiments, their presence
indicates 
that special relativistic time dilation is not rigorously applicable
to particle decays. 
\end{abstract}

\section{ Introduction}
\label{intro}

Unstable particles are a perfect testing
ground for theories that try to unify the principle of relativity with
quantum mechanics 
for (at least) two reasons. First, an unstable particle provides a
very simple 
example of a
non-trivial interacting quantum system. Due to the specific initial condition
characteristic to this problem (there is just one particle in
the initial state), a rigorous description of the decay is possible in a
small Hilbert space that contains only states of the particle and its decay
products, so the solution can be
obtained in a closed form.
  Second, unstable particles played an important role
in confirming predictions of Einstein's special relativity. It was
demonstrated experimentally that the decay of moving particles
slows down in a good agreement with Einstein's time dilation
formula \cite{dilation, muons}.  

For a long time it was believed that relativistic quantum mechanics and
quantum field theory must reproduce exactly the
Einstein's time dilation formula for unstable particles \cite{Exner}. 
However, a detailed quantum relativistic calculation in \cite{mypaper} has shown that there
are small corrections to this formula in the case of
particles with definite momentum. This result was later confirmed in other studies
\cite{Khalfin, Shirokov_decay}. Although surprising, these findings did
not challenge directly the applicability of special relativity to
unstable systems. This is because relativistic transformations connect
results of
measurements of two inertial observers moving with constant velocities
with respect to each other. Then a fair comparison with the time
dilation formula requires consideration of unstable states having
definite values of velocity for both observers. 
Such states do not have a definite
momentum for, at least, one observer. 
Their decay laws were considered in recent article
\cite{Shirokov-osc} which alleges that the decay 
 accelerates
when particle's velocity increases, instead of being slowed down as experiment
shows.  In the present
article we resolve this controversy by performing a detailed 
calculation of the decay laws of 
particles with narrow distributions of velocities observed  from 
moving reference frames. We found that the result of
\cite{Shirokov-osc}
 is based on an incorrect identification of the subspace of
states of the unstable particle in the full Hilbert space of the
system. In sections \ref{ss:general} - \ref{ss:instant}
we present  a rigorous quantum relativistic
framework required for the description of decays. 
 The exact formula for the time dependence of the non-decay
probability in a moving inertial frame of reference is derived 
in section \ref{sc:moving-frame}.
Particular cases of this formula relevant to unstable particles
with sharply defined momenta or velocities
 are considered in sections \ref{sc:mov}  and \ref{sc:general},
respectively. Section \ref{sc:numerical} is devoted to numerical calculations of the
differences between the accurate quantum mechanical result and the standard
time dilation formula (\ref{eq:1b}).
Although these differences are much smaller than the resolution of
modern experiments, their presence is in sharp contradiction with
special relativity which does not tolerate even small deviations from the
Einstein's time dilation.

\section{ Formulation of the problem}
\label{problem}

The decay of unstable particles is described mathematically by the \emph{non-decay
probability}  which has the following definition.
 Suppose that we have a piece of
radioactive material with 
 N unstable nuclei prepared simultaneously at time $t=0$ 
 and denote
$N_u(t)$  the number of nuclei
that
remain undecayed at time $t>0$. Then the non-decay 
probability\footnote{The first argument in $\omega(0,t)$ indicates the
rapidity of the observer that measures the non-decay probability, as described
below.} $\omega(0,t)$ (also called the
\emph{non-decay law} in this paper) is defined as the fraction of
nuclei 
that survived
the decay  in the limit of large $N$

\begin{eqnarray}
\omega(0,t) =\lim_{N \to \infty} N_u(t)/N
\label{eq:dec_law}
\end{eqnarray}

\noindent  So, at each time point the piece of radioactive
material can
be characterized by its \emph{composition}
$(N_u(t)/N; (1 - N_u(t)/N)$,  where
$N_u(t)/N$ is the share of undecayed nuclei in the
sample
and $1 - N_u(t)/N $ is the share of nuclei that
were transformed to the
\emph{decay products}.
In this paper, in the spirit of quantum mechanics, we will treat $N$ 
unstable particles as an ensemble of identically prepared systems
and consider $\omega(0,t)$  as a property of a single particle (nucleus), i.e.,
the probability of finding this particle in the undecayed state.

In this paper we are concerned with comparison of decay
observations made by two observers 
$O$ and $O'$ that move with respect to each other. 
 Without loss of generality we will assume that observer $O'$
moves with respect to $O$ with velocity $(0,0,v)$  along
the $z$-axis and that at time $t=0$ measured by both observer's
clocks the origins of their coordinate systems coincide and all three 
pairs of coordinate axes are parallel ($x \parallel x'$, etc.).
For discussion purposes we will say that $O$ is at rest
while $O'$ is moving. 
 To simplify formulas we will use the \emph{rapidity}
parameter
$\theta = \tanh^{-1}(v/c)$ instead of velocity $v$. 
For example,
in this notation the famous relativistic factor takes the form
$\gamma \equiv (1-v^2/c^2)^{-1/2} = \cosh \theta$. If from the point of view of $O$
the decay is described by the function $\omega(0,t)$, then
the non-decay law of the same particle seen by the moving
observer $O'$ will be denoted by $\omega(\theta,t)$.\footnote{Note
that here parameter $t$ denotes time measured by the clock belonging
to the observer $O'$.} Calculation of
this function and, in particular, the relationship between
$\omega(\theta,t)$ and $\omega(0,t)$ is the major goal of this work.

\section { Postulates of relativity}
\label{ss:postulates}

The purpose of any relativistic theory is to describe the
relationships between measurements made by different 
 \emph{inertial observers} or from different \emph{inertial reference
frames}, 
i.e., reference frames moving in space
with constant velocities along straight lines and without
rotation.
The special \emph{principle of relativity} tells
us 
that these reference frames are exactly equivalent

\begin{itemize}
\item \textbf{Postulate I.}  Experiments identically arranged and
performed in two different inertial reference frames $O$ and $Q$ always yield the
same results.
\end{itemize}

\noindent For any two observers $O$ and $Q$ there is an \emph{inertial
transformation} 
that connects $Q$ to $O$, i.e., the set of rules that allows one to
change from the reference frame $O$ to the reference frame $Q$. 
Each inertial transformation is a combination of space and time translations, 
rotations and boosts. A composition of two inertial transformations is again a
valid inertial transformation. This composition obeys the
associativity law. The inverse of an inertial transformation is also
an inertial transformation. Therefore, they form a
10-parameter Lie group. The structure of this group is specified in
the second postulate

\begin{itemize} 
\item \textbf{Postulate II.} Inertial transformations (space and time
translations, rotations, and boosts) form the \emph{Poincar\'e group}.
\end{itemize}

Obviously, two different observers obtain different results by
measuring observables of the same physical system. Most problems in
physics can be understood as translations of descriptions of the physical system
between different reference frames. For example, if we have a full
description of the system from the point of view of observer $O$, then
the time evolution is obtained by answering the question 
``what is the description of the same system from the point
of view of observer $O''$ that is shifted in time with respect to
$O$?'' 
 It is important to realize that inertial
transformations between observers can be divided into two groups: \emph{kinematical} and
\emph{dynamical}. Kinematical transformations are those whose action
on observables does not depend on the interaction. For example, 
if two stationary observers  look at the unstable particle from
different points in space, they would assign the
same non-decay probability to the particle. The same is true for
stationary observers having different orientations in space. From this
follows\footnote{Our adoption of this postulate means that we are
working in Dirac's \emph{instant form} of relativistic dynamics
\cite{Dirac}. Other forms of dynamics, e.g., the \emph{point form} and
the \emph{front form}, were also introduced by Dirac and are frequently used for
description of relativistic interactions. As discussed in section 
\ref{sc:different}, these forms are not appropriate for the
description of non-decay laws of unstable particles.}

\begin{itemize} 
\item \textbf{Postulate III.} Space
translations and rotations are kinematical.
\end{itemize}

\noindent On the other hand, time translations  produce non-trivial 
changes in the system. 
The composition of the unstable system looks different for observers
$O$ and $O''$ shifted in time with respect to each other.
The exact action of time translations on observables of the
physical system  should be 
obtained as a result of solution of
dynamical equations which depend on the interaction acting in the
system. Hence the following
postulate is true for all types of isolated interacting systems.

\begin{itemize} 
\item \textbf{Postulate IV.} Time translations are dynamical, i.e.,
interaction-dependent.
\end{itemize}

The above postulates I - IV have overwhelming experimental support. They are
assumed to be valid throughout this paper. However they are
not sufficient 
for a full description of the unstable particle in a moving
reference frame. Such a description requires also knowledge of the
nature of boosts. 
Here we have a choice between two paths forward. One
path is to postulate certain properties of boosts. This
path was taken by Einstein. It leads to special relativity and to the
time dilation formula (\ref{eq:1b}) as explained in the next section. In this
paper we will argue in favor of choosing another path: keep postulates
I - IV, add to them well-established postulates of quantum mechanics,
and see what are the implications for the transformations of
observables (in particular, the non-decay probability) with respect to boosts. 
This approach is employed starting from section \ref{ss:general}
throughout the paper.

\section { Particle decay in special relativity}
\label{ss:relativity}

In addition to postulates I - IV,  Einstein's special relativity makes
two more assumptions regarding the nature of boosts

\begin{itemize}
\item \textbf{Assumption V.} Boosts are kinematical. 
\end{itemize}

\noindent By postulating the
kinematical character of boosts, special relativity
insists that the internal composition of a compound system 
 does not depend on the velocity
of the observer. For example, if $O$ and $O'$ are two observers
moving with respect to each other, then,
according to special relativity, both observer will measure the same
composition of the unstable system at time $t=0$. 
 These statements are often considered as self-evident in discussions of
special relativity.  For
example, R. Polishchuk writes in \cite{Polishchuk}
\emph{``Any event that is ``seen'' in one inertial system is ``seen'' in all
others. For example if observer in one system ``sees'' an explosion on
a rocket then so do all other observers.''}
In addition to  the
universality and interaction-independence of boost transformations,
special relativity also postulates the exact transformation laws of physical
observables with respect to boosts. They are
 referred to as \emph{Lorentz transformations}. The most
fundamental are Lorentz transformations 
for space-time coordinates of
events.\footnote{By \emph{event} we understand a measurable physical
process occurring at a certain point in space at one time instant. 
An intersection of trajectories of two point-like
classical particles is an example of such an event. 
Note that in quantum mechanics
the definition of event is problematic as particles do
not have well-defined trajectories. Moreover, it is
known that a particle sharply localized from the point of view of the
observer $O$ loses its localization from the point of view of the moving
observer $O'$ \cite{Newton_Wigner}. So, 
Assumption VI can be applied only in the classical limit. } 

\begin{itemize}
\item \textbf{Assumption VI.} 
 If from the point of view of observer $O$ an event is localized
in a space point $(x,y,z)$ at time $t$, then from the point of view of
observer $O'$ the same event has space-time coordinates $(t', x', y',
z')$  given by Lorentz formulas

\begin{eqnarray}
t' &=& t \cosh\theta - \frac{ z}{c} \sinh\theta,
\label{eq:lorentz-time} \\ 
x' &=& x, \label{eq:lorentz-z} \\
y'&=& y, \label{eq:lorentz-y}\\ 
z' &=& z
\cosh\theta - c t \sinh\theta, \label{eq:lorentz-x}
\end{eqnarray}
\end{itemize}

Let us now demonstrate that in special relativity
the Postulates I - IV and Assumptions V - VI are sufficient to 
unambiguously describe particle decay in
different reference frames without invoking any information about the
interaction governing the decay. Suppose that from the point of 
observer  $O$
the unstable system is prepared in the state with composition  $(1.0; 0.0)$
at rest in 
the origin $x=y=z=0$ at time $t=0$.\footnote{As indicated in the
previous footnote, here we use the classical limit. In quantum
mechanics the  notions ``at rest'' and
``in the origin'' cannot be used simultaneously.} Then observer $O$
may associate the space-time point 

\begin{eqnarray}
(t,x,y,z)_{prep} = (0,0,0,0)
\label{eq:event1}
\end{eqnarray} 

\noindent with  the event of preparation.
In accordance with the dynamical character of time
translations,  the non-decay probability $\omega(0,t)$ decreases with
time. From experiment and quantum mechanical calculations (see, e.g.,
section \ref{sc:numerical}) it is known that 
 the non-decay law  has
an (almost) exponential shape\footnote{The exact form of the
non-decay law is not important for our derivation of eq. (\ref{eq:1b}) here.}

 \begin{eqnarray}
\omega(0,t) 
&\approx& 
\exp(- \frac{ t}{\tau_0}) 
\label{eq:expo}
\end{eqnarray}

\noindent where $\tau_0$ is the \emph{lifetime} of the unstable
particle. At time  $t = \tau_{0}$ 
 the non-decay probability is exactly $e^{-1}$, so that the composition is 
$(0.368; 0.632)$. 
 This ``one lifetime'' event  has 
coordinates 

\begin{eqnarray}
(t,x,y,z)_{life} = (\tau_{0},0,0,0)
\label{eq:event2}
\end{eqnarray}

\noindent according to the observer $O$.

Let us now take the point of view of the moving observer
$O'$. Due to the Assumption V, both
$O$ and
$O'$ agree that at the ``preparation'' and ``lifetime'' events the non-decay
probabilities are 1 and $e^{-1}$, respectively. However, observer $O'$
may not agree with $O$ about
the
space-time coordinates of these events. Substituting (\ref{eq:event1})
 and (\ref{eq:event2}) in (\ref{eq:lorentz-time}) - (\ref{eq:lorentz-x})
we see  that from the point of view of $O'$, the ``preparation'' event  has
coordinates $(0,0,0,0)$, and the ``lifetime'' event  has coordinates
$(\tau_{0}\cosh\theta, 0, 0, - c \tau_0 \sinh\theta)$. Therefore, the time elapsed between
these two events  is $\cosh\theta$ times
longer than in the reference frame $O$. This also means that the decay
law of the particle 
is exactly $\cosh\theta$ slower from the point of view of
the moving observer $O'$. This is reflected in the famous  ``time dilation''
formula

\begin{eqnarray}
\omega(\theta,t) = \omega(0, \frac{t}{\cosh \theta})
\label{eq:1b}
\end{eqnarray}

\noindent This formula  was confirmed in
numerous experiments \cite{dilation}, most accurately for muons accelerated
 to relativistic speeds in a
cyclotron \cite{muons}.
These experimental findings were certainly a triumph of Einstein's theory. 
 However, as we see from the above discussion, in special relativity 
eq. (\ref{eq:1b}) can be derived only under two assumptions V
and VI, which lack proper justification. Therefore, a question
remains whether eq. (\ref{eq:1b}) is a fundamental exact result or simply an
approximation that can be disproved by more accurate measurements?

Our goal in the rest of this paper is to demonstrate that this
classical result of special relativity is not exact. In section
\ref{sc:numerical} we will calculate 
corrections to the formula (\ref{eq:1b}).

\section { Quantum mechanics  of particle decays}
\label{ss:general}

 Let us now turn to the description of an isolated unstable system from the
point of view of relativistic quantum theory. In our approach we will
keep postulates I - IV. However, we are not going to use Assumptions
V and VI.

 We will consider a model theory with  
 particles
$a$, $b$, and $c$, so that particle $a$ is massive 
spinless  and
unstable, 
while its decay
products $b$ and  $c$ are stable and their
 masses satisfy the inequality 

\begin{eqnarray}
m_a > m_b + m_c
\label{eq:masses}
\end{eqnarray}

\noindent which makes 
the  decay $a \to b + c $
energetically possible.
In order to simplify
calculations 
and avoid being
distracted by issues that are not relevant to the problem at hand 
we neglect the spin of the particle
$a$ and assume that there is only one decay mode of this particle, i.e.,
into two decay products $b$ and $c$.
For our discussion, the nature of
the particle $a$ is not that important. For example, 
this could be
a muon or a radioactive nucleus or an atom in an excited state.

When measuring the non-decay law,
experimentalists simply count the number of particles and
determine their types. For example, observations of a muon may result
in only two outcomes. One can find either a non-decayed muon or its
decay products (an electron, a neutrino, and an antineutrino). 
Thus the number of particles is a legitimate
observable,
and for our model system $a \to b+c$  we can introduce 
Hermitian operators
for the number of particles $N_a$, $N_b$, and $N_c$. Since these
observables can be measured simultaneously, we can assume that these
three  operators commute with each other. Two
combinations of their common eigenvalues are allowed in our system:

\begin{eqnarray}
n_a &=& 1, n_b = 0, n_c = 0
\label{eq:eigenval1} \\
n_a &=& 0, n_b = 1, n_c = 1
\label{eq:eigenval2}
\end{eqnarray}

\noindent Hence the Hilbert
space of the unstable system should be represented as a direct sum of two
 orthogonal subspaces\footnote{In principle, a full
description of systems involving these three types of particles must
be formulated in the \emph{Fock space} where integer eigenvalues $n_a, n_b$,
and $ n_c$ are allowed to take any values from zero to infinity.
However, for most unstable particles the interaction between the decay
products in the final state can be ignored, and considering the subspace
(\ref{eq:tensor-product})
of the full Fock space is a reasonable approximation.}

\begin{eqnarray}
\mathcal{H} = \mathcal{H}_ {a} \oplus \mathcal{H}_ {bc}
\label{eq:tensor-product}
\end{eqnarray}

\noindent  where
$\mathcal{H}_ {a}$ is the subspace of states of the unstable particle
$a$ which corresponds to the set of eigenvalues (\ref{eq:eigenval1}),
 and  $\mathcal{H}_ {bc} \equiv \mathcal{H}_ {b} \otimes \mathcal{H}_
{c}$ is the
orthogonal subspace of 
the decay products which corresponds to the set of eigenvalues (\ref{eq:eigenval2}).

We can now introduce a Hermitian operator $T$ 
(also known as ``yes-no experiment'') that corresponds to the
observable ``particle
$a$ exists''.  The operator $T$ can be fully defined by its eigensubspaces and
eigenvalues.
When a measurement performed on the unstable system finds it in a state corresponding to the
particle $a$ (i.e., the state vector is within $\mathcal{H}_a$), 
the value of $T$ is 1. When the decay products
$b+c$ are observed (the state vector lies in $\mathcal{H}_{bc}$), 
the value of $T$ is 0.
  Apparently,  $T$
is a projection operator on the subspace $\mathcal{H}_a$. 
 For each normalized state vector $| \Psi \rangle \in \mathcal{H}$ the probability of
finding the unstable particle $a$  is given by the expectation value of the 
observable  $T$ \cite{Exner}

\begin{eqnarray*}
\omega_{| \Psi\rangle } = \langle \Psi |
T | \Psi \rangle
\end{eqnarray*}

In relativistic quantum mechanics, the dynamics of the
system 
 is described by a 
unitary representation $U_g$ of
the Poincar\'e group in the Hilbert space $\mathcal{H}$ \cite{book}. 
 It is convenient to express 
representatives  $U_g$ as exponential
functions of \emph{generators} $\mathbf{P},
\mathbf{J}, \mathbf{K}$, and $ H$.  Then a general
unitary operator from the set $U_g $ can be always written as a
product

\begin{eqnarray}
U_g = 
e^{-\frac{i}{\hbar}Ht} 
e^{\frac{i}{\hbar}\mathbf{P}\mathbf{a}}
e^{\frac{ic}{\hbar}\mathbf{K} \vec{\theta}}
e^{\frac{i}{\hbar}\mathbf{J} \vec{\phi}}
\label{gen-elem}
\end{eqnarray}

\noindent where $\vec{\phi}$ is the rotation vector, $\mathbf{v} =  c
\frac{\vec{\theta}}{\theta}
\tanh \theta$ is the velocity of the boost, $\mathbf{a}$ is the vector
of space translation, and $t$ is the amount of time translation.
 Generators are Hermitian operators and
correspond to certain \emph{total} observables of the system.
Space translations are generated by the vector of the total
linear momentum  $\mathbf{P}$.
 The generator of rotations $\mathbf{J}$ is the
operator of the total angular momentum. The generator of time
translations $H$ is the Hamiltonian (total energy). The generator of
boosts $\mathbf{K}$ is called the \emph{boost operator}. These
generators 
 must satisfy the 
commutation relations of the Poincar\'e group Lie algebra. 
In this paper we will need, in particular, the following commutators

\begin{equation}
[K_i, P_j] = -i\frac{\hbar}{c^2} H \delta_{ij} 
\label{eq:5.55}
\end{equation}
\begin{equation}
 [K_i, H] = -i\hbar P_i 
\nonumber
\end{equation}

\noindent where $i,j=x,y,z$. They imply the following
useful relationships\footnote{For derivation see section 2.2 in \cite{mybook}.}

\begin{eqnarray}
 e^{\frac{i}{\hbar} \mathbf{K} c \vec{\theta}}
\mathbf{P}  e^{- \frac{i}{\hbar} \mathbf{K} c \vec{\theta}} &=&
\mathbf{P} + \frac{\vec{\theta}}{\theta} [(\mathbf{P} \cdot
\frac{\vec{\theta}}{\theta}) (\cosh \theta - 1) + \frac{1}{c} H \sinh \theta]
\label{eq:6.2} \\ 
e^{ \frac{i}{\hbar} \mathbf{K} c \vec{\theta}} H
e^{ -\frac{i}{\hbar} \mathbf{K} c \vec{\theta}} &=&
 H \cosh \theta + c (\mathbf{P} \cdot
\frac{\vec{\theta}}{\theta}) \sinh \theta
\label{eq:6.3}
\end{eqnarray}

The
 representation (\ref{gen-elem}) allows us to relate
results of measurements in different reference frames.
Let us first take the point of view of the observer $O$ and consider a
vector 
$| \Psi \rangle \equiv |\Psi(0,0)
\rangle \in \mathcal{H}_a$\footnote{In the notation $|\Psi(0,0) \rangle$ the
first argument is the rapidity parameter of the reference frame from
which the state $| \Psi \rangle$ is observed and the second argument
is the time of observation. This is consistent with the convention
adopted for $\omega(\theta,t)$ in section \ref{intro}. } that
describes  a state in which the unstable particle $a$ is found with
100\% certainty. 

\begin{eqnarray*}
\omega_{| \Psi \rangle}(0, 0) 
&=& \langle \Psi| T | \Psi \rangle \nonumber \\
&=& 1
\label{eq:3f}
\end{eqnarray*}
 
\noindent Then the time evolution of the state vector $|\Psi \rangle$ in the
reference frame $O$ is given by

\begin{eqnarray*}
| \Psi (0, t)  \rangle
&=& 
e^{-\frac{i}{\hbar}Ht} | \Psi \rangle
\label{eq:3g}
\end{eqnarray*}

\noindent and the non-decay law  is given by

\begin{eqnarray*}
\omega_{| \Psi \rangle}(0, t) 
&=& \langle \Psi|
e^{\frac{i}{\hbar}Ht} T
e^{-\frac{i}{\hbar}Ht} | \Psi \rangle
\label{eq:3c}
\end{eqnarray*}

\noindent From this equation it is clear that the Hamiltonian $H$
describing the unstable system should not commute with the projection $T$ 

\begin{eqnarray}
 [H, T] \neq 0
\label{eq:44b}
\end{eqnarray}

\noindent Otherwise, the subspace $\mathcal{H}_a$ of states of the
particle $a$ would be invariant with respect to time translations 
and  the particle $a$ would be 
stable ($\omega_{| \Psi \rangle}(0, t) = 1$ for all $t$). 

The moving  observer $O'$ describes the initial state (at $t=0$) 
 by the vector

\begin{eqnarray*}
| \Psi (\theta, 0)  \rangle =  e^{\frac{ic}{\hbar}K_z \theta} | \Psi \rangle
\label{eq:44}
\end{eqnarray*}

\noindent The time
dependence of this state  is

\begin{eqnarray}
| \Psi(\theta, t) \rangle &=&  
e^{-\frac{i}{\hbar}Ht}e^{\frac{ic}{\hbar}K_z \theta} | \Psi \rangle 
\label{eq:44a}
\end{eqnarray}

\noindent Then the non-decay law from the point of view of $O'$ is

\begin{eqnarray}
\omega_{| \Psi \rangle}(\theta, t) 
&=& \langle \Psi(\theta, t) | T | \Psi (\theta,
t) \rangle
\label{eq:3b} \\
&=& \Vert T
| \Psi (\theta, t) \rangle \Vert^2
\label{eq:3d}
\end{eqnarray}

\noindent where the last equation follows from the property $T^2 = T$
of the operator $T$.

\section { Non-interacting representation of the Poincar\'e group}
\label{sc:non-int}

Before calculating the non-decay law in the moving reference frame (\ref{eq:3b}), let us first consider a simpler case when the interaction responsible
for the decay is ``turned off''. Then the 
representation
 of the Poincar\'e group $U_g^0$ acting 
in $\mathcal{H}$ is \emph{non-interacting}.  This representation is constructed
in accordance with the structure of the Hilbert space (\ref{eq:tensor-product}) as

\begin{eqnarray}
U_g^0 \equiv U_g^a \oplus U_g^b
\otimes U_g^c
\label{eq:non-int}
\end{eqnarray}

\noindent  where $U_g^{(a,b,c)}$ are unitary irreducible representations
 of the Poincar\'e group
corresponding to particles $a$, $b$, and $c$, respectively. The
generators 
of the
representation (\ref{eq:non-int}) are denoted by  $\mathbf{P}_0$, $\mathbf{J}_0$, $H_0$, 
and $\mathbf{K}_0$.
The operator of non-interacting \emph{mass}

\begin{eqnarray*}
M_0 = +\frac{1}{c^2} \sqrt{H_0^2 - P_0^2 c^2}
\end{eqnarray*}

\noindent commutes with
 $\mathbf{P}_0$, $\mathbf{J}_0$, $H_0$, 
and $\mathbf{K}_0$.  According to (\ref{eq:masses}), the operator $M_0$ has
a continuous spectrum in the interval $[m_b+m_c, \infty)$ and a
discrete point $m_a$ embedded in this interval.

From definition (\ref{eq:non-int}) it is clear that  the subspaces $\mathcal{H}_ {a}$ and $\mathcal{H}_ {bc}$ are invariant
with respect to   $U_g^0$. 
Moreover, the particle number operator $N_a$ and  the projection 
operator $T$ commute with the non-interacting generators

\begin{eqnarray}
[T,\mathbf{P}_0] = [T,\mathbf{J}_0] = [T,\mathbf{K}_0] =  [T,H_0]=0
\label{eq:T-comm}
\end{eqnarray}

\noindent This implies that the particle $a$ is
stable with respect to time translations and boosts, as expected

\begin{eqnarray*}
\omega_{| \Psi\rangle } (\theta, t) 
&=&  \langle \Psi | e^{-\frac{ic}{\hbar}(K_z)_0  \theta} 
e^{\frac{i}{\hbar}H_0t} T  e^{-\frac{i}{\hbar}H_0t} 
e^{\frac{ic}{\hbar}(K_z)_0  \theta} |\Psi \rangle \label{eq:10} \\
&=&  \langle \Psi |  T  
|\Psi \rangle \nonumber  \\
&=& 1 \nonumber 
\end{eqnarray*}

The primary reason for considering the representation $U_g^0$ is that
it allows us to build 
a convenient basis in the subspace $\mathcal{H}_a$. 
Let us denote $| \mathbf{0} \rangle$ a vector in  $\mathcal{H}_a$ that
corresponds to the particle $a$ with zero momentum. 

\begin{eqnarray*}
\mathbf{P}_0 | \mathbf{0} \rangle &=& \mathbf{0} \\
H_0 | \mathbf{0} \rangle &=& m_ac^2 | \mathbf{0} \rangle
\end{eqnarray*}

\noindent Then we find that the vector  

\begin{eqnarray}
| \mathbf{p} \rangle = 
e^{-\frac{ic}{\hbar}\mathbf{K}_0  \vec{\theta}} 
| \mathbf{0} \rangle
\label{eq:mom-eigen2}
\end{eqnarray}

\noindent  describes the particle $a$ with definite
momentum
$ \mathbf{p}  = \frac{\vec{\theta}}{\theta} m_a c \sinh \theta
$. Indeed,
using 
(\ref{eq:6.2}) we obtain

\begin{eqnarray*}
\mathbf{P}_0 | \mathbf{p}
\rangle 
&=&
\mathbf{P}_0 e^{-\frac{ic}{\hbar}\mathbf{K}_0  \vec{\theta}} | \mathbf{0}
\rangle \\
&=& e^{-\frac{ic}{\hbar}\mathbf{K}_0  \vec{\theta}}
e^{\frac{ic}{\hbar}\mathbf{K}_0  \vec{\theta}}\mathbf{P}_0
 e^{-\frac{ic}{\hbar}\mathbf{K}_0  \vec{\theta}} | \mathbf{0}
\rangle \\
&=& e^{-\frac{ic}{\hbar}\mathbf{K}_0  \vec{\theta}}
(\mathbf{P}_0 + \frac{\vec{\theta}}{\theta} [(\mathbf{P}_0 \cdot
\frac{\vec{\theta}}{\theta}) (\cosh \theta - 1) + \frac{1}{c} H_0 \sinh \theta]) | \mathbf{0}
\rangle \\
&=&  \frac{\vec{\theta}}{\theta} m_ac \sinh \theta 
e^{-\frac{ic}{\hbar}\mathbf{K}_0  \vec{\theta}}
 | \mathbf{0}
\rangle \\
&=& \frac{\vec{\theta}}{\theta}  m_ac \sinh \theta  | \mathbf{p} \rangle 
\end{eqnarray*}

\noindent Since particle $a$ is spinless by our assumption, 
the eigenvectors $| \mathbf{p} \rangle$ of the total momentum operator
$\mathbf{P}_0$
  form a full basis in the subspace $\mathcal{H}_a$,
 so that 

\begin{eqnarray}
\langle \mathbf{p} | \mathbf{p}' \rangle &=& \delta (\mathbf{p}-
\mathbf{p}') \label{eq:wave} \\
T &=& \int d\mathbf{p} |\mathbf{p} \rangle \langle \mathbf{p} |
\label{eq:projector}
\end{eqnarray}

\noindent Then any state $| \Psi \rangle \in \mathcal{H}_a$ of 
the particle $a$   can be represented by a linear combination of these
basis vectors

\begin{eqnarray}
| \Psi \rangle &=& T | \Psi \rangle \nonumber \\
               &=& \int d\mathbf{p}|\mathbf{p}
\rangle \langle \mathbf{p} |\Psi \rangle \nonumber \\
               &=& \int d\mathbf{p} \psi(\mathbf{p})
|\mathbf{p} \rangle
\label{eq:expansion}
\end{eqnarray}

\noindent where
$\psi(\mathbf{p}) = \langle \mathbf{p} | \Psi \rangle$
is the  wave
function  in the momentum representation. In order to ensure the
normalization $\langle \Psi | \Psi \rangle = 1$, the function $\psi
(\mathbf{p})$ must satisfy

\begin{eqnarray}
 \int d\mathbf{p} |\psi (\mathbf{p})|^2 = 1
\label{eq:normalization}
\end{eqnarray}

\noindent Apparently, vectors $|\mathbf{p} \rangle$ themselves are not
normalized. If we want to study the non-decay law
of a state with definite momentum  $\mathbf{p}_0$, we should use a state vector (which
we denote by $|\mathbf{p}_0)$ to distinguish it from $|\mathbf{p}_0
\rangle$) that
 has a
normalized momentum-space
wave function sharply localized near $\mathbf{p}_0$.
In order to satisfy eq. (\ref{eq:normalization}) such a wave function 
may be formally
represented as a square root of the Dirac's delta function
$
\psi(\mathbf{p})=\sqrt{\delta (\mathbf{p} - \mathbf{p}_0)} 
$

The action of boosts on the basis vectors $|\mathbf{p} \rangle$ and
$|\mathbf{p})$ is
known from Wigner's theory of irreducible unitary representations of
the Poincar\'e group (see, e.g., \cite{Wigner, book})

\begin{eqnarray}
  e^{-\frac{ic}{\hbar}(K_z)_0 \theta}|\mathbf{p}
\rangle
&=&  \sqrt{\frac{\Omega_{L
\mathbf{p}}}{\Omega_{\mathbf{p}}}}
|L \mathbf{p} \rangle 
\label{eq:129a} 
\end{eqnarray}

\noindent where

\begin{eqnarray*}
L \mathbf{p} &=& (p_x , p_y,  p_z \cosh \theta + \frac{\Omega_{\mathbf{p}}}{c}
 \sinh
\theta) \\
\Omega_{ \mathbf{p}} &=& \sqrt{m_a^2c^4 + c^2\mathbf{p}^2} 
\label{eq:Omega} 
\end{eqnarray*}

\noindent Using eq. (\ref{eq:129a}) and the property

\begin{eqnarray}
\frac{d\mathbf{p}}{\Omega_{\mathbf{p}}} = 
\frac{d(L\mathbf{p})}{\Omega_{L\mathbf{p}}}
\label{eq:invar}
\end{eqnarray} 

\noindent we can find boost 
transformations for an arbitrary state vector
of the form (\ref{eq:expansion})

\begin{eqnarray*}
  e^{-\frac{ic}{\hbar}(K_z)_0 \theta}|\Psi
\rangle
&=& \int d\mathbf{p} \psi(\mathbf{p})  e^{-\frac{ic}{\hbar}(K_z)_0 \theta}
|\mathbf{p} \rangle \\
&=&\int d\mathbf{p} \psi(\mathbf{p})  \sqrt{\frac{\Omega_{L
\mathbf{p}}}{\Omega_{\mathbf{p}}}}
|L \mathbf{p} \rangle \\
&=&\int d\mathbf{p} \sqrt{\frac{\Omega_{
L^{-1}\mathbf{p}}}{\Omega_{\mathbf{p}}}}
\psi(L^{-1} \mathbf{p})  
| \mathbf{p} \rangle
 \label{eq:129b} 
\end{eqnarray*}

\noindent These transformations can be viewed as
transformations of the corresponding momentum-space wave function, e.g.,

\begin{eqnarray}
e^{-\frac{ic}{\hbar} (K_z)_0
\theta} \psi(\mathbf{p})
&\equiv& \sqrt{\frac{\Omega_{L^{-1}\mathbf{
p }}}{\Omega_{\mathbf{p}}}}
\psi( L^{-1}\mathbf{p})
\label{eq:boost-tr}
\end{eqnarray}

\noindent so that the boost operator can be represented as a differential
operator in the  momentum space

\begin{eqnarray}
 (K_z)_0 \psi(\mathbf{p}) &=& \frac{i \hbar}{c} \lim_{\theta \to 0} \frac{d}{d
\theta}
e^{-\frac{i}{\hbar}(K_z)_0 c\theta}
\psi(\mathbf{p}) \nonumber \\
&=& \frac{i \hbar}{c} \lim_{\theta \to 0} \frac{d}{d \theta}
\sqrt{\frac{\Omega_{L^{-1}\mathbf{
p }}}{\Omega_{\mathbf{p}}}} \psi(p_x, p_y, p_z \cosh
\theta - \frac{\Omega_{\mathbf{p}}}{c}  \sinh \theta) \nonumber \\
&=& -i \hbar ( \frac{\Omega _{ \mathbf{ p}}}{c^2} \frac{d}{dp_z} 
+ \frac{p_z}{2\Omega_{\mathbf{p}}})
 \psi(\mathbf{p})
\label{eq:7.21} 
\end{eqnarray}

For further calculations we will need to define the  
\emph{Newton-Wigner position} operator \cite{Newton_Wigner} in
$\mathcal{H}$ 

\begin{eqnarray*}
\mathbf{R}_0 \equiv 
-\frac{c^2}{2}(H_0^{-1}\mathbf{K}_0 + \mathbf{K}_0 H_0^{-1}) 
\end{eqnarray*}

\noindent which has the property

\begin{eqnarray}
[(R_i)_0, (P_j)_0] &=& i\hbar \delta_{ij}
\label{eq:heis}
\end{eqnarray}

\noindent that can be verified by direct
substitution. 
According to eq. (\ref{eq:7.21})

\begin{eqnarray*}
(R_z)_0 \psi(\mathbf{p})  &=& -\frac{c^2}{2}(H_0^{-1}(K_z)_0 + (K_z)_0 H_0^{-1})
\psi(\mathbf{p}) \nonumber \\
 &=& \frac{i \hbar}{2}(\Omega^{-1} _{ \mathbf{ p}} \Omega _{ \mathbf{
p}} \frac{d}{dp_x} + \Omega _{ \mathbf{p}}\frac{d}{dp_x} 
 \Omega^{-1} _{ \mathbf{ p}} + \frac{p_x c^2}{\Omega^2_{\mathbf{p}}})
\psi(\mathbf{p}) \nonumber \\
&=& i \hbar \frac{d}{dp_x} \psi(\mathbf{p}) 
\label{eq:7.22}
\end{eqnarray*}

\noindent Therefore operator $e^{\frac{ic}{\hbar}(R_0)_z b}$ acts as a
translation operator in the momentum space. In particular, we can
write

\begin{eqnarray*}
  e^{\frac{ic}{\hbar}(R_0)_z m_a c \sinh \theta} \sqrt{\delta(\mathbf{p})}
=  \sqrt{\delta(\mathbf{p} - \mathbf{p}_0)}
\end{eqnarray*}

\noindent where $\mathbf{p}_0 = (0, 0, m_a c \sinh \theta)$.
On the other hand, applying the boost transformation
(\ref{eq:boost-tr}) to the 
momentum eigenfunction we
obtain\footnote{ where $|J| = \frac{\Omega_{L^{-1}
\mathbf{p}}}{\Omega_{\mathbf{p}}}$  is the Jacobian of the
transformation $\mathbf{p} \to L^{-1}
\mathbf{p}$.}

\begin{eqnarray*}
  e^{-\frac{ic}{\hbar}(K_0)_z \theta} \sqrt{\delta(\mathbf{p})}
&=&  \sqrt{\frac{\Omega_{L^{-1}
\mathbf{p}}}{\Omega_{\mathbf{p}}}}
\sqrt{\delta(L^{-1}\mathbf{p})} \\
&=&  \sqrt{\frac{\Omega_{L^{-1}
\mathbf{p}}}{\Omega_{\mathbf{p}}}}
\sqrt{\frac{1}{|J|}\delta(\mathbf{p})} \\
&=&  
\sqrt{\delta(\mathbf{p}- \mathbf{p}_0)} \\
\end{eqnarray*}

\noindent This suggests that momentum eigenvector (\ref{eq:mom-eigen2}) has another
useful representation

\begin{eqnarray}
|\mathbf{p} \rangle
& =&
 e^{\frac{i}{\hbar} \mathbf{R}_0 \cdot \mathbf{p}}|\mathbf{0}
\rangle
\label{eq:alpha}
\end{eqnarray}

\section { Interacting representation of the Poincar\'e group}
\label{ss:instant}

Let us now ``turn on'' the interaction responsible for the decay and
discuss 
the interacting representation $U_g$ of the
Poincar\'e group in $\mathcal{H}$ with generators $\mathbf{P}$,
$\mathbf{J}$, $\mathbf{K}$, and $H$. According to our postulates III
and IV, the generators of space translations and rotations are
interaction-free,

\begin{eqnarray*}
\mathbf{P}&=&\mathbf{P}_0 \\
\mathbf{J} &=& \mathbf{J}_0 
\end{eqnarray*}

\noindent while the generator of time translations (the Hamiltonian $H$)
contains an interaction-dependent term $V$.

\begin{eqnarray*}
 H &=& H_0+V
\end{eqnarray*}

\noindent From the commutator (\ref{eq:5.55}) it then follows \cite{Dirac} that the 
generators of boosts must be interaction-dependent as well

\begin{eqnarray*}
 \mathbf{K} &=& \mathbf{K}_0+\mathbf{W} 
\end{eqnarray*}

\noindent where $\mathbf{W} \neq 0$.
This means that we are working in  the so-called Dirac's
 \emph{instant form of dynamics}.  We will further assume that 
the interacting representation $U_g$
belongs to the Bakamjian-Thomas form of
dynamics \cite{Bakamjian_Thomas}, which is characterized by the
property
 that the interacting operator of mass $M \equiv
c^{-2}\sqrt{H^2 - \mathbf{P}_0^2 c^2}$ commutes with the
Newton-Wigner position operator\footnote{The possibilities for the
interaction to be not in the Bakamjian-Thomas instant form are
discussed in section \ref{sc:different}.}

\begin{eqnarray}
[\mathbf{R}_0, M] = 0
\label{eq:nw-comm}
\end{eqnarray}

Our next goal is to define the basis of common  eigenvectors of commuting operators $\mathbf{P}_0$
and $M$ in $\mathcal{H}$.\footnote{In addition 
to these two operators, whose eigenvalues are used for
labeling eigenvectors $| \mathbf{p},m \rangle$, there are other
independent operators 
in the
mutually commuting set containing $\mathbf{P}_0$ and $M$. These are, for example,
the operators of the square of the total angular momentum $\mathbf{J}_0^2$
and the projection of the total angular momentum on the $z$-axis
$(J_0)_z$. Therefore a unique characterization of any basis vector requires
specification of all corresponding quantum numbers as $| \mathbf{p},m , j^2, j_z, \ldots
\rangle$. However these quantum numbers  are not relevant for our discussion
and we omit them.} These eigenvectors must satisfy conditions

\begin{eqnarray}
\mathbf{P}_0 
| \mathbf{p},m \rangle 
&=&   
 \mathbf{p} | \mathbf{p},m \rangle 
\label{eq:mom-eigen} \\
 M| \mathbf{p}, m \rangle &=&   m| \mathbf{p}, m \rangle
\label{eq:mass-eigen}
\end{eqnarray}

\noindent They are also eigenvectors of the interacting
Hamiltonian $H = \sqrt{M^2 c^4 + \mathbf{P}_0^2 c^2}$ 

\begin{eqnarray*}
 H| \mathbf{p}, m \rangle &=&  
\omega_{ \mathbf{p}}| \mathbf{p}, m \rangle 
\end{eqnarray*}

\noindent where $\omega_{ \mathbf{p}} \equiv \sqrt{m^2c^4 +
c^2\mathbf{p}^2}$.\footnote{Note the difference between $\omega_{
\mathbf{p}}$ that depends on the eigenvalue $m$ of the interacting
mass operator and $\Omega_{
\mathbf{p}}$ in eq. (\ref{eq:Omega}) that depends on the fixed value 
of mass $m_a$ of the particle $a$.}
In the zero-momentum eigensubspace of the momentum operator $\mathbf{P}_0$
we can
introduce a basis $| \mathbf{0},m \rangle$ of   eigenvectors of the
 interacting mass $M$ 

\begin{eqnarray*}
\mathbf{P}_0 | \mathbf{0}, m \rangle &=& \mathbf{0} \\
M | \mathbf{0}, m \rangle &=& m| \mathbf{0}, m
\rangle 
\end{eqnarray*}

\noindent Then the basis $| \mathbf{p}, m \rangle$ in the entire Hilbert space $\mathcal{H}$ can be built
by formula (cf. eq. (\ref{eq:mom-eigen2}))

\begin{eqnarray*}
| \mathbf{p}, m \rangle = 
e^{-\frac{ic}{\hbar}\mathbf{K}  \vec{\theta}} 
| \mathbf{0}, m \rangle
\end{eqnarray*}

\noindent where $\mathbf{p} = mc \vec{\theta} \theta^{-1} \sinh \theta$. 
These eigenvectors are normalized to delta functions

\begin{eqnarray}
\langle \mathbf{q},m | \mathbf{p}, m' \rangle &=& \delta (\mathbf{q}-
\mathbf{p}) \delta (m - m')
\label{eq:28a} 
\end{eqnarray}

\noindent The
actions of inertial transformations on these states are well-known \cite{book}.
In particular, for  boosts along the
$z$-axis (cf.  eq. (\ref{eq:129a})) and  time translations we
obtain

\begin{eqnarray}
  e^{-\frac{ic}{\hbar}K_z \theta}|\mathbf{p}, m
\rangle
&=&  \sqrt{\frac{\omega_{\Lambda
\mathbf{p}}}{\omega_{\mathbf{p}}}}
|\Lambda \mathbf{p},m \rangle \label{eq:29a} \\
e^{\frac{i}{\hbar}Ht} |\mathbf{p}, m
\rangle &=&   e^{\frac{i}{\hbar} 
\omega_{ \mathbf{p}}t} | \mathbf{p},m \rangle
\label{eq:29b}
\end{eqnarray}

\noindent where

\begin{eqnarray*}
\Lambda \mathbf{p}&=& (p_x , p_y,  p_z \cosh \theta + \frac{\omega_{\mathbf{p}}}{c}
 \sinh
\theta) \label{eq:lambdaz}
\end{eqnarray*}

\noindent Next we notice  that due to eqs. (\ref{eq:heis}) and (\ref{eq:nw-comm}) vectors $e^{\frac{i}{\hbar} \mathbf{R}_0 \cdot \mathbf{p}}|\mathbf{0}
, m \rangle$ also satisfy eigenvector equations (\ref{eq:mom-eigen}) -
(\ref{eq:mass-eigen}), so they must be proportional to the basis
vectors $|\mathbf{p}, m \rangle$

\begin{eqnarray*}
|\mathbf{p}, m \rangle =
\gamma(\mathbf{p},m) e^{\frac{i}{\hbar} \mathbf{R}_0 \cdot \mathbf{p}}|\mathbf{0}
, m \rangle
\end{eqnarray*}

\noindent where $\gamma(\mathbf{p},m)$ is an unimodular
factor. Unlike in (\ref{eq:alpha}), we cannot conclude that
$\gamma(\mathbf{p},m) = 1$, because, generally, the action of $e^{\frac{i}{\hbar}
\mathbf{R}_0 \cdot \mathbf{p}}$ on eigenvectors $|\mathbf{q}, m \rangle$ involves
multiplication by a unimodular scalar in addition to the shift of momentum. However, 
if the 
interaction is not pathological we can assume that the factor
$\gamma(\mathbf{p},m)$ is smooth, 
i.e., without rapid oscillations. This property will be used in
derivation of eq. (\ref{eq:appr_dec}).

 Obviously,  vector  $| \mathbf{0}
\rangle$ can be expressed as a linear combination of 
 zero-momentum basis vectors
$| \mathbf{0}, 
m \rangle$,
so we can write\footnote{We will assume that interaction responsible
for the decay  does not
change the spectrum of mass. 
In particular, we will neglect the possibility of
existence of bound states of particles $b$ and $c$, i.e., 
discrete eigenvalues of $M$ below $m_b+m_c$. Then the spectrum of $M$ (similar
to the spectrum of $M_0$) is continuous in the
interval $[m_b+m_c, \infty)$, and
integration in (\ref{eq:integral}) should be performed from
$m_b + m_c$ to infinity.}

\begin{eqnarray}
|\mathbf{0} \rangle &=&  
\int \limits_{m_b + m_c}^{\infty} dm \mu(m) | \mathbf{0},m \rangle
\label{eq:integral}
\end{eqnarray}

\begin{figure}
\epsfig {file=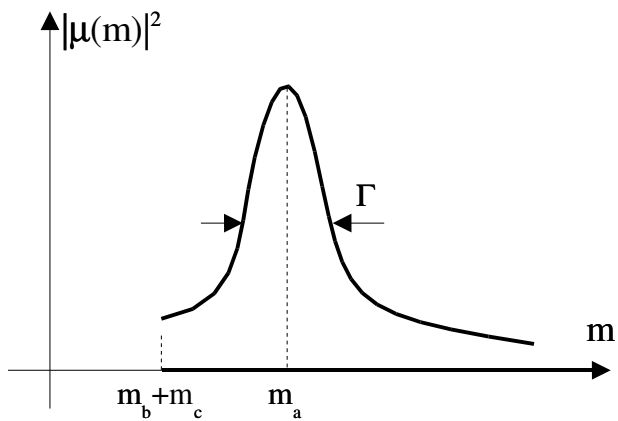}
\caption{ Mass distribution of a typical unstable particle.}
\end{figure}

\noindent where function $ |\mu(m) |^2 $ describes the \emph{mass
distribution} 
 of the
unstable particle. For most realistic unstable systems the mass distribution
$|\mu(m)|^2$ has the Breit-Wigner form (see Fig. 1)\footnote{Strictly
speaking, the center of the resonance (\ref{eq:63}) could be different
from the mass $m_a$ that particle $a$ has in the absence of decay
interaction. However, we will disregard this possibility here.}

\begin{eqnarray}
|\mu(m)|^2 &\approx& \frac{\alpha \Gamma /2 \pi}{\Gamma^2/4  + (m - m_a )^2}, \mbox{
  } if \mbox{ } m \geq m_b + m_c 
\label{eq:63}\\
|\mu(m)|^2 &=& 0, \mbox{
  } if \mbox{ }m < m_b + m_c 
\label{eq:64}
\end{eqnarray}

\noindent where  parameter $\alpha$ is a
normalization 
factor required to ensure that the mass distribution (\ref{eq:63}) - (\ref{eq:64}) is normalized
to unity.

We now use eqs. (\ref{eq:alpha}) and (\ref{eq:integral}) to expand 
the vector 
$|\mathbf{p} \rangle$ in the basis $| \mathbf{p},m \rangle$

\begin{eqnarray}
|\mathbf{p} \rangle
&=&  e^{\frac{i}{\hbar} \mathbf{R}_0 \mathbf{p}}
\int \limits_{m_b + m_c}^{\infty} dm \mu(m) | \mathbf{0},m
\rangle \nonumber \\
  &=&  
\int \limits_{m_b + m_c}^{\infty} dm \mu(m) \gamma(\mathbf{p}, m) | \mathbf{p},m
\rangle 
\label{eq:p-expansion}
\end{eqnarray}

\noindent
Then, from (\ref{eq:28a}) we obtain a useful formula

\begin{eqnarray}
\langle \mathbf{q} |\mathbf{p},m \rangle &=& 
 \int \limits_{m_b + m_c}^{\infty} dm' \mu^*(m') \gamma^*(\mathbf{q}, m') \langle \mathbf{q},m'
|\mathbf{p},m \rangle \nonumber \\
&=& 
 \gamma^*(\mathbf{p}, m) \mu^*(m) \delta(\mathbf{q-p}) 
\label{eq:35a}
\end{eqnarray}

\section { General formula for the non-decay law}
\label{sc:moving-frame}

Suppose that vector $| \Psi \rangle$ in  (\ref{eq:expansion})
describes a state of the unstable particle $a$ from the point of view of
the observer $O$. The time
dependence of this state seen from the moving reference frame $O'$ is obtained by applying eqs (\ref{eq:44a}),  
(\ref{eq:29a}), (\ref{eq:29b}), and (\ref{eq:p-expansion})

\begin{eqnarray*}
| \Psi(\theta, t) \rangle 
& =& \int
 d\mathbf{p}  \psi (\mathbf{p})
e^{-\frac{i}{\hbar}Ht} e^{\frac{ic}{\hbar}K_z \theta}  
 |\mathbf{p}\rangle \\
& =& \int
 d\mathbf{p}  \psi (\mathbf{p}) \int \limits_{m_b + m_c}^{\infty} dm \mu(m) 
\gamma(\mathbf{p}, m) e^{-\frac{i}{\hbar}Ht} e^{\frac{ic}{\hbar}K_z \theta}  
 |\mathbf{p},m \rangle \\
& =& \int
 d\mathbf{p}  \psi (\mathbf{p})
\int \limits_{m_b + m_c}^{\infty} dm \mu(m) \gamma(\mathbf{p}, m)
e^{-\frac{i}{\hbar}\omega_{\Lambda^{-1}
\mathbf{p}}t} \sqrt{\frac{\omega_{\Lambda^{-1} \mathbf{p}}}{\omega_{\mathbf{p}}}}  
 |\Lambda^{-1}\mathbf{p},m \rangle
\end{eqnarray*}

\noindent The inner product of this vector with $| \mathbf{q} \rangle $ is
found by using (\ref{eq:35a})

\begin{eqnarray*}
&\mbox{ }& \langle \mathbf{q}  
|  \Psi(\theta, t)
\rangle \nonumber 
\\
&=&  \int 
d\mathbf{p}  \psi (\mathbf{p})
\int \limits_{m_b + m_c}^{\infty} dm \mu(m)\gamma(\mathbf{p}, m) 
e^{-\frac{i}{\hbar}\omega_{\Lambda^{-1}
\mathbf{p}}t} 
\langle \mathbf{q}  |\Lambda^{-1} \mathbf{p} ,m \rangle 
\sqrt{\frac{\omega_{\Lambda^{-1} \mathbf{p}}}{\omega_{\mathbf{p}}}} \nonumber \\ 
&=& \int  d\mathbf{p} \psi (\mathbf{p})  \int \limits_{m_b + m_c}^{\infty} dm  |\mu(m)|^2
 \gamma(\mathbf{p}, m)\gamma^*(\Lambda^{-1} \mathbf{p}, m) 
e^{-\frac{i}{\hbar}\omega_{\Lambda^{-1} \mathbf{p}}t} 
\delta(\mathbf{q} - \Lambda^{-1} \mathbf{p}) 
\sqrt{\frac{\omega_{\Lambda^{-1} \mathbf{p}}}{\omega_{\mathbf{p}}}}
\label{eq:36a} 
\end{eqnarray*}

\noindent Introducing new integration variables
$
\mathbf{r} = \Lambda^{-1} \mathbf{p} 
$
and taking into account (\ref{eq:invar}), this equation can be rewritten as 

\begin{eqnarray*}
&\mbox{ }& \langle \mathbf{q}  
|  \Psi(\theta, t)
\rangle \nonumber \\
&=& 
\int \limits_{m_b + m_c}^{\infty} dm \int 
 d \mathbf{r} 
\frac{\omega_{\Lambda \mathbf{r}}}
{\omega_{\mathbf{r}}}
\sqrt{\frac{\omega_{\mathbf{r}}}{\omega_{\Lambda \mathbf{r}}}} \psi(\Lambda \mathbf{r})
\gamma(\Lambda \mathbf{r}) \gamma^*( \mathbf{r}) \vert \mu(m)\vert^2 
e^{-\frac{i}{\hbar}\omega_{ \mathbf{r}}t} 
    \delta(\mathbf{q} - \mathbf{r}) \nonumber \\
&=&  
\int \limits_{m_b + m_c}^{\infty} dm  
\sqrt{\frac{\omega_{\Lambda \mathbf{q}}}
{\omega_{\mathbf{q}}}} 
\psi(\Lambda \mathbf{q}) \gamma(\Lambda \mathbf{q}, m) 
\gamma^*( \mathbf{q}, m)  \vert \mu(m)\vert^2 
e^{-\frac{i}{\hbar}\omega_{ \mathbf{q}}t}   
\end{eqnarray*}

\noindent
 The non-decay probability in the reference frame $O'$ is then found 
by substituting
(\ref{eq:projector}) 
in
eq. (\ref{eq:3b})

\begin{eqnarray}
\omega_{|\Psi \rangle}(\theta, t) &=& \int  d\mathbf{q} \langle
\Psi(\theta, t)  
 |\mathbf{q} \rangle\langle \mathbf{q}   | \Psi(\theta, t)\rangle
\nonumber \\
 &=& \int  d\mathbf{q} |\langle \mathbf{q}  
  |\Psi (\theta, t)\rangle\ |^2  \nonumber \\
&=& \int  d\mathbf{q} \vert 
\int \limits_{m_b + m_c}^{\infty} dm  
\sqrt{\frac{\omega_{\Lambda \mathbf{q}}}
{\omega_{\mathbf{q}}}} 
\psi(\Lambda \mathbf{q}) \gamma(\Lambda \mathbf{q}, m) 
\gamma^*( \mathbf{q}, m)  \vert \mu(m)\vert^2 
e^{-\frac{i}{\hbar}\omega_{ \mathbf{q}}t}    \vert^2
\label{eq:50}
\end{eqnarray} 

\noindent which is an exact formula valid for all values of $\theta$ and
$t$.

\section { Decay law in the reference frame at rest}
\label{sc:mov} 

In the reference frame at rest ($\theta = 0$),  
formula (\ref{eq:50}) simplifies

\begin{eqnarray*}
\omega_{ | \Psi \rangle}(0, t) 
= \int d \mathbf{q}  |\psi(\mathbf{q})|^2
|\int \limits_{m_b + m_c}^{\infty} dm |\mu(m)|^2 
  e^{-\frac{i}{\hbar} \omega_{\mathbf{q}} t}|^2
\end{eqnarray*}

\noindent Let us consider a particular case of this expression when the
unstable particle has a
well-defined momentum, i.e., described by the state vector $|
\mathbf{p})$. 
If the particle
$a$ is in the state $|\mathbf{0})$ with
 zero momentum then $\psi(\mathbf{q}) =
\sqrt{\delta(\mathbf{q})}$ and from the above equation we obtain

\begin{eqnarray}
\omega_{ |\mathbf{0})}(0,t) 
&=& 
|\int \limits_{m_b + m_c}^{\infty}dm |\mu(m)|^2 
  e^{-\frac{i}{\hbar} mc^2 t}|^2
\label{eq:34}
\end{eqnarray}

\noindent which is the standard formula for the non-decay law of a particle at
rest.\footnote{By noting that for the particle at rest its energy is
 identified with $mc^2$, this formula can be compared, for example, 
 with eq. (3.8)
in \cite{Fonda}.} In particular, if we substitute (\ref{eq:63}) - (\ref{eq:64}) for the mass distribution
$|\mu(m)|^2$ we
obtain approximate exponential decay
(\ref{eq:expo})
where the lifetime is given by 

\begin{eqnarray}
\tau_0 = \hbar/(\Gamma c^2)
\label{eq:lifetime}
\end{eqnarray}

\noindent If the particle $a$ has a definite non-zero momentum $\mathbf{p}$,
then 

\begin{eqnarray}
\psi(\mathbf{q}) =
\sqrt{\delta(\mathbf{ q-p})}
\label{eq:sqrt}
\end{eqnarray}

\noindent and

\begin{eqnarray}
\omega_{ |\mathbf{p})}(0,t) 
&=& 
|\int \limits_{m_b + m_c}^{\infty}dm |\mu(m)|^2 
  e^{-\frac{i}{\hbar}\omega_{\mathbf{p}} t}|^2
\label{eq:35}
\end{eqnarray}

\noindent In a number of works \cite{mypaper, Khalfin, Shirokov_decay} it was noticed that if one
interprets the state $|\mathbf{p})$ (where $|\mathbf{p}| = m_a c \sinh
\theta$) as a state of unstable particle
moving with definite speed $|\mathbf{v}| = c \tanh \theta$, then the decay
law of the moving particle (\ref{eq:35}) cannot be connected with the non-decay law
of the particle at rest (\ref{eq:34}) by Einstein's formula (\ref{eq:1b}), i.e.,

\begin{eqnarray}
\omega_{ |\mathbf{p})}(0,t) 
&\neq& 
\omega_{ |\mathbf{0})}(0, t/ \cosh \theta)
\label{eq:35b}
\end{eqnarray}

\noindent This observation
prompted authors of \cite{mypaper, Khalfin, Shirokov_decay} to question the applicability of Einstein's
special
relativity to particle decays.
However, at a closer inspection it appears that this result 
does not  contradict
Einstein's time dilation formula (\ref{eq:1b}) directly. Formula
(\ref{eq:1b}) 
refers to
observations made on the
same particle from two frames of reference moving with respect to each
other. If from the point of view of observer $O$ the particle is
described by the state vector $|\mathbf{0}) $ which has
zero momentum and zero velocity, then from the point of view of
$O'$ this particle is described by the state 

\begin{eqnarray}
| \mathbf{p}) = e^{\frac{ic}{\hbar}\mathbf{K}  \vec{\theta}} |
\mathbf{0})
\label{eq:velo-state}
\end{eqnarray} 

\noindent which 
 is an eigenstate of
the velocity operator\footnote{Indeed, 
taking into account  
$V_z | \mathbf{0}) \approx \mathbf{0}$ and eqs. (\ref{eq:6.2}) - (\ref{eq:6.3}), we obtain

\begin{eqnarray*}
V_z e^{\frac{ic}{\hbar}K_z  \theta} | \mathbf{0}) 
&=& e^{\frac{ic}{\hbar}K_z  \theta}
e^{-\frac{ic}{\hbar}K_z  \theta} V_z e^{\frac{ic}{\hbar}K_z  \theta} | \mathbf{0}
) \nonumber \\
&=& e^{\frac{ic}{\hbar}K_z  \theta}
\frac{V_z - c \tanh \theta}{ 1 - \frac{V_z \tanh \theta}{c} } | \mathbf{0}
) \nonumber \\
&\approx&  -c \tanh \theta e^{\frac{ic}{\hbar}K_z \theta} | \mathbf{0}
)
\label{eq:velocity-eig}
\end{eqnarray*}}

\begin{eqnarray}
 \mathbf{V} = c^2\mathbf{P}_0/H
\label{eq:velocity}
\end{eqnarray}

\noindent but is not an eigenstate of the
momentum operator $\mathbf{P}_0$. Therefore, strictly speaking, its non-decay law is not described by $\omega_{ |\mathbf{p})}(0,t)$.
In order to
compare with Einstein's formula (\ref{eq:1b}), we need to calculate the decay
law in the moving frame of reference $\omega_{ | \Psi \rangle}(\theta, t) $.
This is done in 
section \ref{sc:general}.

\section {Decays caused by boosts}
\label{sc:dec-boost}

Let us now discuss the non-decay probability $\omega_{ | \Psi
\rangle}(\theta, 0) $ at $t=0$ in the moving
frame of reference. Instead of eq. (\ref{eq:50}), it is more
convenient to use the general definition (\ref{eq:3d}) which expresses
$\omega_{|\Psi \rangle}(\theta, 0)$ as the square of the norm of the
projection of $|\Psi (\theta, 0) \rangle$ on the subspace $\mathcal{H}_a$. 
We are going to prove that for a normalized $|\Psi \rangle$ this 
probability cannot be equal to 1 for all non-zero $\theta$. 
 Suppose that this statement is wrong, so that (in Agreement with
Assumption V) for any 
$|\Psi \rangle \in \mathcal{H}_a$ and any $\theta > 0$, the vector 
$e^{\frac{ic}{\hbar}K_z \theta} | \Psi \rangle$ belongs
to $\mathcal{H}_a$. Then the subspace $\mathcal{H}_a$  is invariant
under action of boosts $e^{\frac{ic}{\hbar}K_z \theta}$
and
 operator $K_z$ 
commutes with the projection $T$. Then from the Poincar\'e commutator
(\ref{eq:5.55}) and $[T, (P_0)_z]= 0$ it follows by Jacobi identity that

\begin{eqnarray*}
[T,H] &=& \frac{ic^2}{\hbar}[T,[K_z,(P_0)_z]] \\ &=&  \frac{ic^2}{\hbar} [K_z,[T,
(P_0)_z]] - \frac{ic^2}{\hbar}[(P_0)_z,[T,K_z]] \\ &=& 0
\end{eqnarray*}

\noindent which contradicts eq. (\ref{eq:44b}). 
This contradiction implies that the state $e^{\frac{ic}{\hbar}K_z \theta} | \Psi \rangle$
does not correspond to the particle
$a$ with 100\% probability; this state must contain contributions from the
decay products  even at
$t=0$

\begin{eqnarray}
e^{\frac{ic}{\hbar}K_z \theta} | \Psi \rangle
&\notin& \mathcal{H}_a
\label{eq:not-in} \\
\omega_{| \Psi \rangle} (\theta, 0) &<& 1, \mbox{  } for \; \; \theta \neq 0
\label{eq:not-in2}
\end{eqnarray}

\noindent This is the ``decay caused by boost''.
Thus we found that Assumption V from section \ref{ss:relativity}
is not correct. Sometimes this assumption is  formulated as
 \cite{Giunti} 
\emph{``Flavor is the quantum number that
distinguishes the different types of quarks and leptons. It is a
Lorentz invariant quantity. For example, an electron is seen as an
electron by any observer, never as a muon.''}
Although the above statement
about the electron is correct (because the electron is a stable particle),
this is not true about the muon which, according to (\ref{eq:not-in}) can be seen as a single particle
by the observer at rest and as a group of three decay products (an electron, a
neutrino, and an antineutrino)
by the moving observer.

\section {Decay law in the moving reference frame} 
\label{sc:general}

 Unfortunately exact
evaluation of eq. (\ref{eq:50}) for $\theta \neq 0$ is not possible, so we need to make
approximations.
Let us discuss  properties of the initial state $|\Psi \rangle$ in
more detail.
First, in all realistic cases this state is not an eigenstate of the
total momentum operator, so the wave function is not localized at one
point in the momentum space (as assumed in (\ref{eq:sqrt}))  but has a
spread 
(or uncertainty) of
momentum $|\Delta \mathbf{p}|$ and, correspondingly an uncertainty of position $|\Delta
\mathbf{r}| \approx \hbar/ |\Delta \mathbf{p}|$. 
On the other hand, the state $|\Psi
\rangle \in \mathcal{H}_a$ is not an eigenstate of the mass operator $M$, and $|\Psi \rangle$ is
characterized by the uncertainty of mass $\Gamma$ (see Fig. 1) that is
related to the
lifetime of the particle $ \tau_0$ by formula (\ref{eq:lifetime}). 
It is important to note that in all cases of practical interest the
mentioned quantities are related by  inequalities

 \begin{eqnarray}
|\Delta \mathbf{p}| &\gg& \Gamma c 
\label{eq:gg}\\
|\Delta \mathbf{r}| &\ll&  c \tau_0
\label{eq:gg2}
\end{eqnarray}

\noindent In particular, the latter inequality means that the uncertainty of position is mush less than the
distance passed by light during the lifetime of the
particle. 
For example, in the case of muon $\tau_0 \approx 2.2 \cdot
10^{-6}$s and, according to (\ref{eq:gg2}), the spread of the wave function in the position space must
be much less than 600m, which is a reasonable assumption.
 Therefore, we can safely assume that 
 the factor  $|\mu(m)|^2$ in (\ref{eq:50})
has a sharp peak near the value $m = m_a$. Then we can move  the value of
the smooth  function $\sqrt{\frac{\omega_{\Lambda \mathbf{q}}}
{\omega_{\mathbf{q}}}}
\psi(\Lambda \mathbf{q}) \gamma(\Lambda \mathbf{q},m)
\gamma^*(\mathbf{q},m)$  
at $m=m_a$ outside 
the integral by $m$.\footnote{Note that if we 
assumed instead of
(\ref{eq:gg}) that the spread of the momentum-space
wave function is much less than $\Gamma c$ we could have moved the
value of the smooth function $\vert \mu(m_0)\vert^2$ outside the
integral by $m$. This would mean that the non-decay law in the moving
frame of reference is controlled
by the spread of the particle wavefunction in the momentum space
rather than by its mass uncertainty which disagrees with experimental observations.}

\begin{eqnarray}
\omega_{| \Psi \rangle}(\theta, t)
&\approx&  \int  d\mathbf{q} 
\vert \sqrt{\frac{\Omega_{\Lambda \mathbf{q}}}
{\Omega_{\mathbf{q}}}} 
\psi(\Lambda \mathbf{q}) \gamma(\Lambda \mathbf{q}, m_a) 
\gamma^*( \mathbf{q}, m_a) \vert ^2   \vert 
\int \limits_{m_b + m_c}^{\infty} dm  
  \vert \mu(m)\vert^2 
e^{-\frac{i}{\hbar}\omega_{ \mathbf{q}}t}    \vert^2 \nonumber \\ 
&=& \int  d\mathbf{q}
\frac{\Omega_{L \mathbf{q}}}
{\Omega_{ \mathbf{q}}}
|\psi(L \mathbf{q}) |^2
\vert 
\int \limits_{m_b + m_c}^{\infty} dm  
 \vert \mu(m)\vert^2 
e^{-\frac{i}{\hbar}\omega_{ \mathbf{q}}t}    \vert^2 \nonumber \\
&=& \int  d\mathbf{p}
|\psi( \mathbf{p})|^2 
\vert 
\int \limits_{m_b + m_c}^{\infty} dm  
 \vert \mu(m)\vert^2 
e^{-\frac{i}{\hbar}\omega_{ L^{-1} \mathbf{p}}t}    \vert^2 
\label{eq:appr_dec}
\end{eqnarray}

\noindent Now we are going to use formula (\ref{eq:appr_dec}) 
to calculate the decay
law  for an approximate eigenstate of momentum in $\mathcal{H}_a$
whose wave function $\psi( \mathbf{p})$ is localized near zero
momentum $\mathbf{p} = \mathbf{0}$, though inequality (\ref{eq:gg}) still remains
valid. We denote this state vector by symbol
$|\mathbf{0}]$. The state $e^{\frac{ic}{\hbar}K_z  \theta} | \mathbf{0}] $ 
is an approximate eigenstate of the velocity
operator (\ref{eq:velocity}) for all values of $\theta$.\footnote{See footnote 17.} 
Therefore $\omega_{| \mathbf{0}]}(\theta, t)$ in eq. (\ref{eq:omega-0-t}) can be
understood as the non-decay law of a particle with definite speed 
$v = c \tanh \theta$.

 Note that the second factor
in the integrand in (\ref{eq:appr_dec}) is a slowly varying function of $\mathbf{p}$. 
Therefore, we can
set $|\psi( \mathbf{p})|^2 \approx  \delta(\mathbf{p})$
in eq. (\ref{eq:appr_dec}) and obtain

\begin{eqnarray}
\omega_{|\mathbf{0}]}(\theta, t) 
&\approx&  
\vert 
\int \limits_{m_b + m_c}^{\infty} dm  
 \vert \mu(m)\vert^2 
e^{-\frac{it}{\hbar}\sqrt{m^2 c^4 + m_a^2 c^4\sinh^2 \theta} }
\vert^2 
\label{eq:omega-0-t}
\end{eqnarray}

\noindent If we approximately\footnote{See remark after eq. (\ref{eq:velo-state}).} identify $m_a c \sinh \theta$ with the
momentum $|\mathbf{p}|$ of the particle $a$ from the point of view of
the moving
observer $O'$ then

\begin{eqnarray}
\omega_{|\mathbf{0}]}(\theta, t) 
&\approx& 
|\int \limits_{m_b + m_c}^{\infty} dm |\mu(m)|^2 
  e^{-\frac{i}{\hbar}\omega_{\mathbf{p}} t}|^2
\label{eq:62}
\end{eqnarray}

\noindent So, in this approximation the non-decay law (\ref{eq:62}) 
in the frame of reference $O'$ moving with the
speed $c \tanh \theta$ takes the same form
as the non-decay law (\ref{eq:35}) of a particle moving with momentum $m_a
c
\sinh \theta$ with respect to the stationary observer $O$.

\section { Numerical results}
\label{sc:numerical}

In this section we are going to perform numerical calculations of the
differences between the
classical Einstein's formula (\ref{eq:1b})

\begin{eqnarray}
\omega_{|\mathbf{0}]}^{class}(\theta, t) = \omega_{|\mathbf{0}]}(0,
\frac{t}{\cosh \theta}) 
\label{eq:time-dilation}
\end{eqnarray}

\noindent and the actual non-decay law (\ref{eq:omega-0-t}) of a moving particle
having either definite momentum or sharply defined speed. 
In these calculations we assumed that the mass distribution
$|\mu(m)|^2$ of the unstable system (see eqs. (\ref{eq:63}) -
(\ref{eq:64}) and  Fig. 1) is centered at the value of
mass $m_a= 1000$ MeV/$c^2$, the total mass of the decay
products
 was $m_b + m_c = 900$ MeV/$c^2$,
and the width of the mass distribution was $\Gamma$= 20 MeV/$c^2$.
 These values do not
correspond to any real particle, but they are typical 
for strongly decaying
baryons.

It is
convenient to measure time in units of the classical lifetime $\tau _0
\cosh \theta$.
 Denoting $\chi \equiv t/(\tau _0 \cosh \theta)$, we find that
 Einsteinian non-decay laws (\ref{eq:time-dilation}) for
any value of $\theta$ are
given by the same universal  function
$\omega^{class}(\chi)$.  This function
was calculated 
 for values of $\chi$ in the interval
from 0 to 6
 with the step of 0.1   The calculations were performed by direct numerical integration
of eq. (\ref{eq:34}) using the \emph{Mathematica} program shown 
below

\begin{verbatim}
gamma = 20

mass = 1000

theta = 0.0

Do[Print[(1/0.9375349) Abs[NIntegrate [gamma/(2 Pi) / (gamma^2/4
+(x - mass)^2) Exp[ I t Sqrt [x^2 + mass^2 (Sinh [theta])^2]
Cosh [theta] / gamma], {x, 900, 1010, 1100, 300000}, MinRecursion -> 3,
MaxRecursion -> 16, PrecisionGoal -> 8, WorkingPrecision -> 18]]^2],
{t, 0.0, 6.0, 0.1}]
\end{verbatim}

\begin{figure}
\epsfig {file=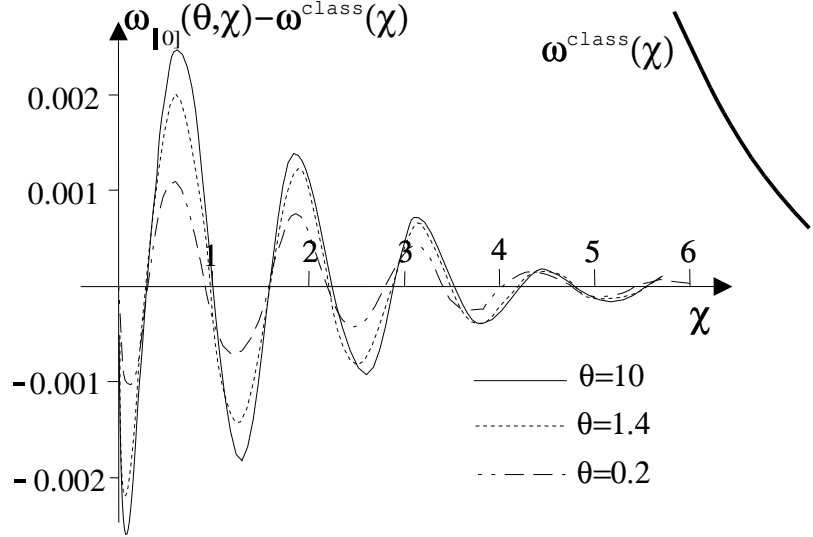}
\caption{ Corrections to the Einstein's ``time dilation'' formula
(\ref{eq:1b}) for the non-decay law of unstable particle 
 moving with the speed $v = c \tanh \theta$.
Parameter $\chi$ is time measured in units of $\tau_0 / \cosh \theta$.}
\end{figure}

\noindent As expected, function $\omega^{class}(\chi)$ is very 
close to the exponent
$e^{- \chi}$.\footnote{This function is represented by a thick solid line in
Fig. 2. The magnitudes of $\omega^{class}(\chi)$ for small
values of the argument $\chi$ are too large to be shown on the scale
of Fig. 2.} Next we used eq. (\ref{eq:omega-0-t}) and the above 
\emph{Mathematica} program to calculate the
non-decay laws $\omega_{|\mathbf{0}]}(\theta, \chi)  $ in  moving
reference frames for three values
 of the parameter $\theta$ (=\verb/theta/), namely 0.2, 1.4, and
10.0, that  correspond
to velocities of 0.197c, 0.885c, and 0.999999995c, respectively.
These calculations qualitatively confirmed the time dilation formula
(\ref{eq:time-dilation}) to the accuracy of better than 0.3\%. However, they also
revealed important differences 
$ \omega_{|\mathbf{0}]}(\theta, \chi)  - \omega ^{class}(\chi) $
which are plotted as thin lines  in Fig. 2.\footnote{Results
presented in Fig. 2 are different from those in Fig. 1 in ref.
\cite{mypaper} due to the more accurate calculation procedure employed
in the present work.}

The lifetime of the particle $a$ considered in our example 
($ \tau_0  \approx 2 \times 10^{-22}$  s) is 
too short to be observed experimentally. So, calculated corrections to
the Einstein's time dilation law have only illustrative
value. However, from these data we can estimate the magnitude of
corrections for particles whose time-dependent non-decay laws can be
measured in a laboratory, e.g., for muons ($\tau_0 \approx 2
\times 10^{-6}$s, $\Gamma \approx 2 \times 10^{-9} eV/c^2$, $m_a \approx
105 MeV/c^2$). 
Taking into account that the magnitude of corrections is
roughly proportional to the ratio $\Gamma/m_a$ \cite{mypaper,
Shirokov_decay},
 we can
expect that for  muons the maximum correction should be about $2 \times
10^{-18}$ which is much smaller than the accuracy
of modern experiments (about $10^{-3}$ \cite{muons}). So, experimental observation
of the predicted corrections requires significant improvements of existing 
experimental techniques.

Note that formulas (\ref{eq:omega-0-t}) and
(\ref{eq:62}) for the particle seen from the moving frame are
approximate   in the sense that they ignore the
``decay caused by boost'' expressed by eq. (\ref{eq:not-in2}).
Nevertheless, our major approximation (\ref{eq:gg}) is well justified and it cannot
explain the discrepancy of our results (Einsteinian time dilation of the
decay plus small interaction-dependent corrections) with the
conclusion of ref.
 \cite{Shirokov-osc} about the acceleration of the decay of moving particles. 
 In our view this  conclusion does not refer to the experimentally
measured non-decay law that is defined as the probability of finding
one unstable particle. Instead of eq. (\ref{eq:3b}), the
non-decay 
probability was defined
in \cite{Shirokov-osc} by formula

\begin{eqnarray*}
\omega_{|\Psi \rangle}(\theta, t) 
&=& 
|\langle \Psi(\theta, 0)| e^{-\frac{i}{\hbar} Ht} | \Psi(\theta, 0)
\rangle|^2
\label{eq:shirok}
\end{eqnarray*}

\noindent whose physical meaning remains unclear.

\section {Particle decays in different Dirac's forms of dynamics}
\label{sc:different}

In this article we assumed that the interaction responsible for
the particle decay has the Bakamjian-Thomas instant form of
dynamics. This assumption was used to simplify calculations, but there
is no good reason to believe that real interactions have this form. Then, naturally, one may ask a question:
``Is there another form of dynamics in which Einstein's time dilation
formula (\ref{eq:1b}) is exactly true?'' The answer is \emph{No}. 
In any instant form of dynamics (including non-Bakamjian-Thomas
instant forms
of dynamics) the boost operators contain interaction terms, so the results (\ref{eq:not-in}) - (\ref{eq:not-in2}) are
still valid, and eq.  (\ref{eq:1b}) holds only
approximately. 
In the point form dynamics\footnote{ where generators of boosts
$\mathbf{K}_0$ and
rotations $\mathbf{J}_0$ are kinematical, while generators of space
translations
$\mathbf{P}$ and time
translations $H$ contain interaction terms.}  the subspace
$\mathcal{H}_a$ of the unstable particle is invariant with respect to boosts,
so there can be no boost-induced decays
(\ref{eq:not-in}). However, due to the interaction-dependence of the total momentum
operator $\mathbf{P}$, one should expect decays induced
by space translations

\begin{eqnarray}
e^{\frac{i}{\hbar}P_z a} | \Psi \rangle
&\notin& \mathcal{H}_a, \mbox{  } for \; \; a\neq 0
\label{eq:not-in-tr} 
\end{eqnarray}

\noindent  This
prediction  is not confirmed by
experiments which show that the composition of an unstable
particle 
is not affected by space translations of the observer.\footnote{
An even more striking contradiction between predictions of the point
form dynamics and observations is that the decay of moving particles
\emph{accelerates} $\cosh \theta$ times instead of experimentally
observed slowdown \cite{mypaper}.}
Therefore, the point form of dynamics is not acceptable for the
description of decays. Similarly, never observed translation- and/or rotation-induced
decays are characteristic for all non-instant forms of
dynamics, e.g., the front form. Therefore only the instant form of dynamics is appropriate
for the description of particle decays. Note, however, that the decay slowdown and
the absence of translation-induced decays are firmly established only
for unstable systems whose
lifetimes are long enough to be measured experimentally, i.e., those
governed by electromagnetic and weak interactions. For strongly interacting
resonances these properties
are beyond experimental resolution, and the non-instant forms of dynamics
cannot be ruled out.

\section {Summary}
\label{sc:discussion}

In this paper we analyzed the relationships between the non-decay laws in the moving
reference frame and in the reference frame at rest. We used a rigorous
quantum relativistic approach that is applicable to any unstable
system independent on the nature of interaction governing the
decay.  A complete description of dynamics in different reference
frames was obtained by using relativistic Postulates I - VI and rules of quantum
mechanics only. We found that Assumptions V and VI of special relativity are
not needed. Moreover, their consequences (the universal and exact time
dilation of the decay of moving particles) are in contradiction with
rigorous calculations.  Although the time dilation  (\ref{eq:1b}) of special relativity
was qualitatively confirmed by our
results, we also found small corrections to this formula that
depend on the strength of interaction. 
 In a broader sense our results indicate that clocks viewed from the moving
reference frame do not go exactly $\cosh \theta$ slower. The exact
amount of time dilation depends on the physical makeup of the clock
and on interactions responsible for the operation of the clock.

I am grateful to Dr. M. I. Shirokov for reading the
manuscript,
valuable comments, and discussions.

\end{document}